\documentclass[twocolumn, superscriptaddress, amsmath, secnumarabic, floatfix, amssymb, nobibnotes, aps, pra, reprint]{revtex4-1}
\usepackage{nicefrac}
\usepackage{amsfonts}
\usepackage{amsmath}
\usepackage{graphicx}
\usepackage{comment}
\usepackage[caption=false]{subfig}
\usepackage[utf8]{inputenc}
\usepackage{float}
\usepackage{color}
\usepackage[showzone=false]{datetime2}
\usepackage{multirow}
\usepackage{url}
\usepackage{ulem}

\def\bra #1{\langle #1\vert}
\def\ket #1{\vert #1\rangle}
\def\braket #1#2{\langle #1 \vert #2\rangle}
\def\ketbra #1#2{\vert #1\rangle \langle #2\vert}
\def\kettbra#1{\ketbra{#1}{#1}}

\def\Hdmin #1 #2{H_{\text{min}}(#1|#2)}

\def\pGue #1{p_{\text{guess}}\left( #1 \right)}
\def\prN #1 {p_{#1}}
\def\sigN #1 {\sigma_{#1}^2}
\def\state #1{\rho_{#1}}
\DeclareMathOperator{\tr}{tr}

\begin{document}

\title{Simple source device-independent continuous-variable\\ quantum random number generator}
\author{P. R. Smith}%
\affiliation{Toshiba Research Europe Ltd, 208 Cambridge Science Park, Milton Road, Cambridge, CB4 0GZ, United Kingdom}
\affiliation{Cambridge University Engineering Department, 9 JJ Thomson Avenue, Cambridge, CB3 0FA, United Kingdom}
\author{D. G. Marangon}%
\email{Corresponding author: davide.marangon@crl.toshiba.co.uk}
\affiliation{Toshiba Research Europe Ltd, 208 Cambridge Science Park, Milton Road, Cambridge, CB4 0GZ, United Kingdom}
\author{M. Lucamarini}%
\affiliation{Toshiba Research Europe Ltd, 208 Cambridge Science Park, Milton Road, Cambridge, CB4 0GZ, United Kingdom}
\author{Z. L. Yuan}%
\affiliation{Toshiba Research Europe Ltd, 208 Cambridge Science Park, Milton Road, Cambridge, CB4 0GZ, United Kingdom}
\author{A. J. Shields}%
\affiliation{Toshiba Research Europe Ltd, 208 Cambridge Science Park, Milton Road, Cambridge, CB4 0GZ, United Kingdom}

\begin{abstract}
\noindent Phase-randomized optical homodyne detection is a well-known technique for performing quantum state tomography.
So far, it has been mainly considered a sophisticated tool for laboratory experiments but unsuitable for practical applications.
In this work, we change the perspective and employ this technique to set up a practical continuous-variable quantum random number generator.
We exploit a phase-randomized local oscillator realized with a gain-switched laser to bound the min-entropy and extract true randomness from a completely uncharacterized input, potentially controlled by a malicious adversary.
Our proof-of-principle implementation achieves an equivalent rate of 270 Mbit/s.
In contrast to other \textsl{source-device-independent} quantum random number generators, the one presented herein does not require additional active optical components, thus representing a viable solution for future compact, modulator-free, certified generators of randomness.
\end{abstract}

\maketitle

\section{Introduction}

\noindent Randomness is an essential resource in many areas of science and information technology.
The problem of accessing true randomness has recently led to the proposal of a variety of random number generator designs \cite{herrero2017quantum}.
So-called ``device-independent'' (DI) quantum-random-number generators (QRNGs) minimize the assumptions underlying the randomness generation process by associating it
with the violation of Bell inequalities \cite{pironio2010random,liu2018high,liu2018device,Shen2018}. However, the complexity of the setups and small generation rates strongly limit their practical use.

Trusted QRNGs exploit a trusted environment for the preparation and the measurement of the quantum states from which the random numbers are extracted.
This makes it possible to build compact and fast generators, suitable for real-world applications.
However, due to their very nature, any hidden side channel in the trusted environment compromises the unpredictability of the generated numbers.

Semi-device-independent QRNGs represent an intermediate solution to achieve a high level of practicality.
 They introduce a minimal set of assumptions either on the measurement \cite{Lunghi2015,Cao2016,Brask2017,VanHimbeeck2017} or on the preparation \cite{Fiorentino2007,Li2011,Vallone2014} parts of the generator.
The latter, so-called \textsl{source} device-independent (SDI) QRNGs, relieve the user (Alice) from the burden of a perfect quantum state preparation.
The most paranoid scenario is when an evil party (Eve) replaces Alice's input state with her own state so that the generated numbers look random to Alice but actually are not.
In this framework, Alice can
counteract Eve's attack by applying measurements that are out of Eve's reach.

In this work we introduce a continuous-variable (CV) SDI QRNG with which we demonstrate generation rates of 270 Mbit/s.
Typical CV-QRNGs feature optical homodyne detection to measure a quadrature observable of an input quantum state \cite{Gabriel2010,Symul2011,Haw2015,Shi2016,Haylock2018,guo2018enhancing,Raffaelli2018,Zheng2018,Gehring2019}.
The quadrature is selected by the phase of a classical field, the so-called local oscillator (LO), which interferes with the input field.
The LO is typically a continuous-wave laser.
In our SDI protocol, the laser is pulsed and gain switched such that each pulse features a random phase \cite{Jofre2011,abellan2014ultra,yuan2014robust,mitchell2015,abellan2015}.
This allows us to use the tomographic technique of phase-randomized homodyne detection \cite{Munroe1995,Leonhardt1996,Lvovsky2001} for random number generation, the security of which follows from randomly changing the phase of the LO.

\begin{center}
\begin{figure*}[h]
\includegraphics[width=\textwidth]{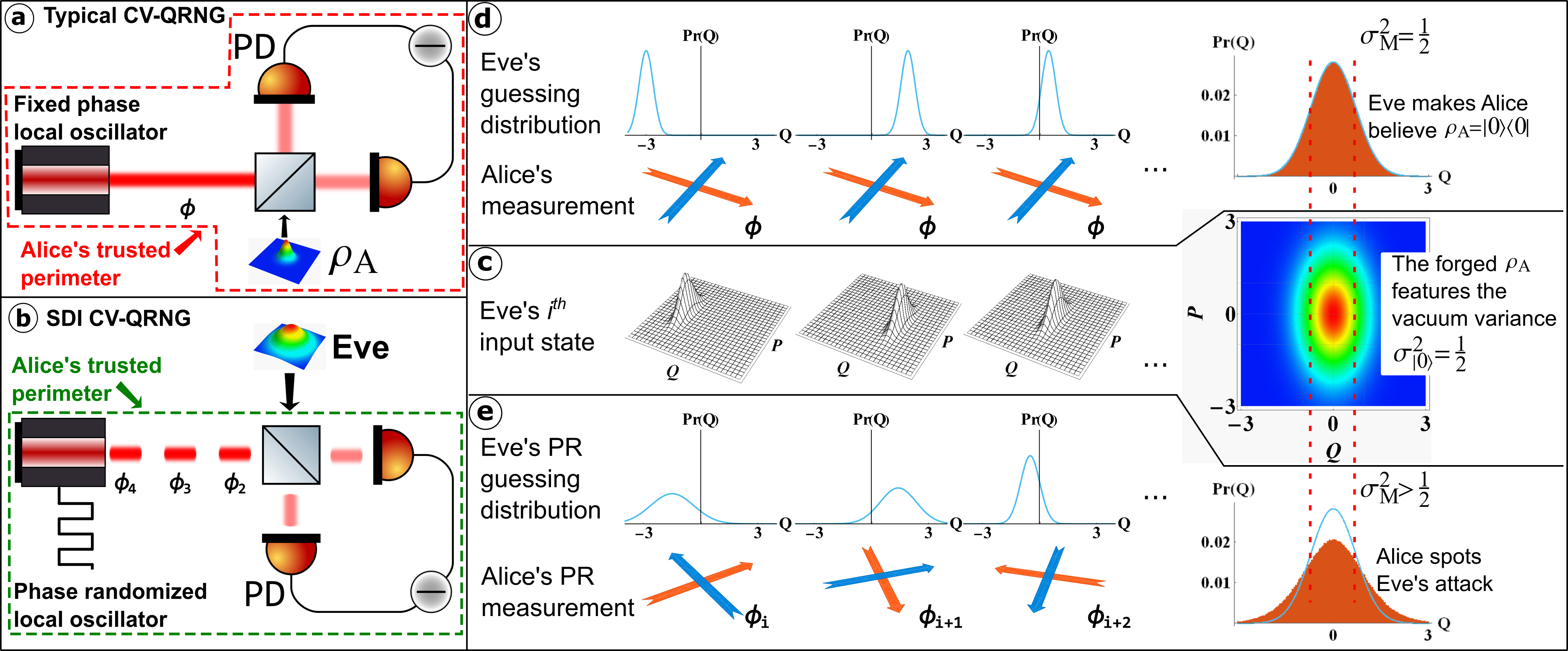}
 \caption{\textbf{(a)} Schematics of a typical CV-QRNG. The input state $\rho_A$ is assumed to be prepared by Alice, so it is trusted and lies within the security perimeter (red dashed line). The LO has a fixed phase, letting Alice measure one specific quadrature of the input field.  \textbf{(b)} Schematics of SDI CV-QRNG. The input state is untrusted and can even be prepared by Eve, so it lies outside the security perimeter (green dashed line). The  LO is phase randomized by using a gain switched laser, which allows Alice to measure random quadratures of the input field.
\textbf{(c)} Example of attack to \textbf{(d)} a typical CV-QRNG and \textbf{(e)} an SDI CV-QRNG, if Eve controls the input state.
In \textbf{(d)}, the LO has a fixed phase. Eve forges $\rho_A$ using Q-displaced  squeezed states and guesses the raw numbers with high probability.
However, Alice thinks she is measuring the vacuum state because the decomposition chosen by Eve mimics the Q distribution of the vacuum. This compromises the security of the system.
In \textbf{(e)}, Alice does not trust the input state as she is in the SDI setting. She uses a phase-randomized LO so that Eve's guessing probability depends on the random angle of the quadrature selected by the LO. Alice can then spot the attack because the measurement distribution is wider than the one she expected from the vacuum input state. }
 \label{fig_big_1}
\end{figure*}
\end{center}

Unlike other recently introduced SDI CV-QRNGs~\cite{Marangon2017,Xu2017,Avesani2018}, ours features the same optical setup as a typical CV-QRNG. No additional optical components are  required. 
The phase randomization of the LO, which is the key element of our generator, is obtained without resorting to a phase modulator. This let us relax the security assumptions on the input state without increasing the complexity of the setup. 
We refer to Fig.~\ref{fig_big_1} to illustrate the difference between our SDI CV-QRNG and a typical one.

CV-QRNGs use balanced homodyne detection (BHD) to measure a quadrature observable $\mathcal{Q}$ of an input state $\rho_A$.
This corresponds to Alice applying the quadrature operator $\hat{Q}_{\theta}= \frac{1}{\sqrt{2}}\left( e^{i\frac\theta2} \hat{a}^\dag+e^{-i\frac\theta2} \hat{a} \right)$ on $\rho_A$, where $\hat{a}^{\dag}$ and $\hat{a}$ are the creation and annihilation operators such that $[\hat{a},\hat{a}^\dag]=1$ holds and $\theta$ is the phase of the LO, which is usually \textsl{fixed}.
The eigenvalue equation for $\hat{Q}_{\theta}$ is $\hat{Q}_{\theta}\ket{q_{\theta}} =q_\theta \ket{q_{\theta}}$, with $q_\theta$ a real number. 

Since the generator is characterized by a finite resolution $\delta$, the measurements of the quadratures return the raw random numbers $q_{\theta,k}$, where  $k$ is the bin index of the intervals $I_{\delta}^k=\left(k-\frac{\delta}{2},k+\frac{\delta}{2}\right]$, with the central bin corresponding to $k=0$ \footnote{Typically the number of bins matches the cardinality of the ADC alphabet, see \cite{Haw2015}.}.
\onecolumngrid
\twocolumngrid The discretized quadrature spectrum, $Q_{\theta,\delta}$ defines the random variable associated with the measurement outcomes: each result is obtained with probability $\mathfrak{p}(q_{\theta,k})=\tr\left[\rho_A\hat{Q}^k_{\theta,\delta}\right]=\int_{I_{\delta}^k} dq\bra{q_\theta}{\rho_A}\ket{q_\theta}$, where 
$\hat{Q}^k_{\theta,\delta}=\int_{I_{\delta}^k} dq\ket{q_\theta}\bra{q_\theta}$ are the elements of Alice's positive operator-valued measure (POVMs) applied on $\rho_A$.
If the input state can be trusted to be pure, the maximal number of independent and identically distributed (iid) bits extractable per  measurement is given by the min-entropy $H_{\text{min}} \left( Q_{\theta,\delta} \right) = -\log_2 \pGue {Q_{\theta,\delta}}$, where $\pGue {Q_{\theta,\delta}} = \max_k\mathfrak{p}(q_{\theta,k})$ is the guessing probability \cite{konig2009operational}
Typically CV-QRNGs trust the input state to be the vacuum \cite{Gabriel2010,Symul2011,Haw2015,Shi2016,Haylock2018,guo2018enhancing,Raffaelli2018,Zheng2018}, $\rho_A=\ket{0}\bra{0}$ [see Fig.~\ref{fig_big_1}-a], for which the LO's phase is irrelevant due its to the rotational invariance in phase space.
The associated outcome distribution $|\braket 0 {q}|^2$ is Gaussian with zero mean and variance $\sigma^2_{\ket{0}}=1/2$, such that the min-entropy is given by
\begin{equation}\label{bound_vacuum}
H_{\text{min}} \left( Q_{\delta}\right) _{\ket 0} =-\log_2 \text{erf}\left(\frac{\delta}{2}\right)\,.
\end{equation}
However, in the SDI paradigm,  the measurement is assumed to be under Alice's control whereas the input state is uncharacterized and even assumed to be controlled by Eve [see Fig.~\ref{fig_big_1}(b)].

An example attack [Fig. \ref{fig_big_1}(c).] can clarify the difference between the two cases [Figs.~\ref{fig_big_1}(d) and \ref{fig_big_1}(e)].
Suppose that Eve controls the input state.
In the non-SDI case, Fig.~\ref{fig_big_1}(d), she knows that Alice measures $\rho_A$ along the $\mathcal{Q}$ quadrature selected by the LO phase $\theta$, which is fixed.
Eve can then input a displaced squeezed state such that she can predict $q_{\theta,k}$ with high confidence.
To conceal her attack, Eve displaces the states so that the probabilities $\mathfrak{p}(q_{\theta,k})$ measured by Alice are the same as those she would expect from her trusted input vacuum state.
Clearly, Alice could never spot this attack and she would \textsl{overestimate} the actual randomness of the samples.
In the limit of infinite squeezing, Eve could predict each outcome with certainty and the actual min-entropy would become zero.
In the SDI case on the contrary, Fig.~\ref{fig_big_1}(e),

Alice measures the input field on a quadrature randomly selected by the LO, which is assumed to be inaccessible to Eve.
This foils Eve's strategy based on a squeezed input.
Without knowing Alice's LO phase, Eve cannot determine the correct squeezing direction for her attack.
This makes the distribution measured by Alice broader than the one corresponding to the vacuum, $\sigma^2_M > \sigma^2_{\ket{0}}$, which unveils the attack.

\section{Bound for the entropy\\ with phase randomization}
In the presence of an adversary controlling the source, the maximal number of iid bits distillable with a randomness extractor is given by the min-entropy $H_{\text{min}} \left( Q_{\theta,\delta}|E\right)$ conditioned on the quantum side information available to Eve.
This quantity considers a purification $\rho_{AE}$ of the input state $\rho_A$:
the system $E$, e.g. a quantum memory, is entangled with Alice's system $A$ and held by Eve who measures it to predict $Q_{\theta,\delta}$.
The quantum conditional min-entropy is then defined as
\begin{equation}\label{PGuess1}
H_{\text{min}} \left( Q_{\theta,\delta}|E\right)=-\log_2 \max_{\{\hat{Q}_{\theta,E}\}} \sum_{k=-\infty}^{\infty} \mathfrak{p}(q_{\theta,k}) \tr\left[ \hat{Q}_{\theta,E}^k \rho_E^{k} \right]
\end{equation}
with $\rho_E^{k}$ being the post-Alice-measurement state of $E$, on which Eve applies the POVM $\{\hat{Q}_{\theta,E}\}$ \cite{furrer2014position,coles2017entropic}.

In the following we will lower bound $H_{\text{min}} \left( Q_{\theta,\delta}|E\right)$ by phase randomizing Alice's states, a procedure typically used to enhance the performance of quantum key distribution with weak coherent states \cite{Gottesman2004, Lo2005a}.

To show the efficacy of this procedure, consider the following example. Eve shares with Alice a two-mode squeezed-vacuum state
$\state {AE} = (1-\gamma^2)\sum_{n,m=0}^{\infty}\gamma^{m+n} \ket{n}_E\bra{n}\otimes\ket{m}_A\bra{m} \,,$
where $\gamma = \tanh{r}$ and $r$ the squeezing parameter.

Although the quadrature fluctuations look random to Alice, the numbers are not private, as Eve can learn them from her part of the state.
However, if Alice's input is phase randomized, $\rho_{AE}$ becomes
$
\state {AE,\varphi_{av}}^{\text{pr}} = (1-\gamma^2)\sum_{n=0}^{\infty}\gamma^{2n} \ket{n}_E\bra{n}\otimes\ket{n}_A\bra{n} \,,
$

which is a separable state that guarantees the privacy of Alice's numbers.

We generalize this example by considering the density matrix of a pure bipartite state in the Fock basis
\begin{equation}\label{SharedState}
\state {AE} = \sum_{k,l,n,m} \rho^{k,l}_{n,m}\ket{k}_E\bra{l} \otimes \ket{n}_A\bra{m}.
\end{equation}
Alice phase randomizes the input by applying the phase shift operator $\hat{U}_{\varphi} = e^{-i\varphi \hat{n}}$ to her part of the system,
\begin{align}\label{PhaseRandomized:1}
\state {AE}^{\text{pr}} =& \sum_{k,l,n,m} \rho^{k,l}_{n,m}\ket{k}_E\bra{l}\otimes\hat{U}_{\varphi}\ket{n}_A\bra{m}\hat{U}^{\dag}_\varphi \nonumber\\
=& \sum_{k,l,n,m} \rho^{k,l}_{n,m}\ket{k}_E\bra{l}\otimes\ket{n}_A\bra{m}e^{-i(n-m)\varphi}, \end{align}
with the phase uniformly distributed in the interval $\varphi \in \{0,2\pi\}$.
Since Eve does not know the $\varphi$ values, the state $\state {AE}^{\text{pr}}$ is averaged to
\begin{equation}\label{PhaseRandomized:2}
\state {AE, \varphi_{av}}^{\text{pr}} = \sum_{k,l,n}\rho^{k,l}_{n,n}\ket{k}_E\bra{l}\otimes\ket{n}_A\bra{n} \,.
\end{equation}
This relation shows that phase randomization returns the same outcome as a quantum non demolition measurement of the photon number \cite{zhao2010security} that disentangles $A$ from $E$. In fact, Eq. (\ref{PhaseRandomized:2}) can be also rewritten in a manifestly separable form \cite{pirandola2013entanglement}.

Equation~\eqref{PhaseRandomized:2} also entails that Alice's most generic input state after phase randomization is a classical mixture of Fock states, as is clear from $\tr_E \state {AE,\varphi_{av}}^{\text{pr}} = \sum_{n} p_n\ket{n}_A\bra{n}_A$
with $p_n= \sum_k \rho^{k,k}_{n,n}$. Therefore  
it is equally secure to consider that Eve inputs such a mixture rather than preparing a general state $\state {AE}$.
The side information is now related to the ensembles $\{p_n,\ket{n}\}$ 
and the conditional min-entropy becomes
\begin{equation}\label{PGuess2}
H_{\text{min}} \left( Q_{\delta}|E\right)_{\text{pr}}=-\log_2\max_{\{p_n,\ket{n}\}} \sum_{n} p_n \max_{k}\tr\left[\hat{Q}^k_{\delta}\ket{n}\bra{n}\right]
\end{equation}
with the external maximization performed over all Eve's possible $\{p_n,\ket{n}\}$ compatible with $\state {A,\varphi_{av}}^{\text{pr}}$ \cite{Law2014}.

Alice can now easily bound Eq. (\ref{PGuess2}) by noticing that the largest guessing probability is obtained when Eve inputs the vacuum state $\ket{0}\bra{0}$.
In fact, the argument of the external maximization is a convex combination of probabilities; hence it is automatically upper bounded by its maximum element, that is, $\pGue {Q_{\delta}}_{\ket 0}\simeq {\delta}/{ \sqrt{\pi}}$.
The vacuum is the Fock state with the narrowest uncertainty in the phase space, which implies
\begin{equation}\label{PGuess3}
 \max_k \tr {\hat{Q}^k_{\delta} \kettbra n } < \max_k \tr {\hat{Q}^k_{\delta} \kettbra 0 }=\pGue Q_{\ket 0}
 \end{equation}
 for $n\geq 1$.
Hence, among all the possible $\{p_n,\ket{n}\}$, the trivial decomposition $\{p_0=1,\ket{0}\}$ is the best forging strategy for Eve, which implies the following bound for the conditional min-entropy
\begin{equation}\label{Entropy-PR-Bound}
H_{\text{min}} \left( Q_{\delta}|E\right)_{\text{pr}} \geq H_{\text{min}}\left( Q _{\delta}\right)_{\ket 0} \,.
\end{equation}
Consequently, when Alice performs phase randomization, Eve's best attack is to input the vacuum state.

\section{SDI CV-QRNG with phase randomized LO}

The scheme presented in the previous section is SDI if we assume that a phase modulator randomizing the input state is part of Alice's measuring setup and Eve cannot access it.
This assumption is hardly justifiable in practice.
For example, this phase modulator could be probed by external bright pulses \cite{lucamarini2015practical}.
Fortunately, there is no need for this phase randomizer in our setup as the phase randomization comes for free from a LO generated by a gain-switched laser.

As we show in Appendix A, Eve's density matrix after Alice's state phase randomization and  quadrature measurement with a fixed phase $\theta$,
\begin{equation}\label{first_equivalence}
\state E^{\text{I}} = \tr_A \left[\left(\text{Id}_E\otimes\hat{Q}_{\theta,\delta}^k\right)^{\dagger} \int^{2\pi}_0 \frac{d\varphi}{2\pi} \hat{U}^{\dag}_{\varphi} \state {AE} \hat{U}_\varphi \left(\text{Id}_E\otimes\hat{Q}_{\theta,\delta}^k\right) \right ],
\end{equation}
 is equal to the phase averaged matrix obtained by Alice after applying a randomly $\varphi$-phase shifted  quadrature operator $\hat{Q}^{\text{ps}}_{\theta,\phi}$,
\begin{equation}\label{second_equivalence}
\state E^{\text{II}} = \tr_A \left[\int^{2\pi}_0 \frac{d\varphi}{2\pi}  \left(\text{Id}_E\otimes \hat{Q}^{\text{ps}}_{\theta,\phi}\right) \state {AE} \left(\text{Id}_E\otimes \hat{Q}^{\text{ps}}_{\theta,\phi} \right)^\dagger \right],
\end{equation}
where $\hat{Q}^{\text{ps}}_{\theta,\phi}= \hat{U}_{\varphi}\hat{Q}_{\theta,\delta}^k\hat{U}^{\dag}_\varphi$.
Therefore the two situations are equivalent securitywise.

The feasibility of the SDI protocol is greatly simplified by having $\state E^{\text{I}} = \state E^{\text{II}} $ in Eqs. (\ref{first_equivalence}) and (\ref{second_equivalence}).
Firstly, because applying $\hat{U}_{\varphi}=e^{-i\varphi \hat{n}}$ to $\hat{Q}_{\theta,\delta}^k$ corresponds to shifting the LO by a phase $\varphi$,
we can replace the phase modulator with a phase-randomized LO, by exploiting the process of phase diffusion in gain switched lasers \cite{abellan2014ultra,henry1982theory}.
This has practical consequences on security as Eve cannot tamper with a phase modulator placed on the input port.
Moreover if a real phase modulator were used to randomize the LO phase, another RNG would be necessary to properly drive it.

\section{Experimental realization}

\begin{figure}
 \centering
 \includegraphics[width=0.49\textwidth]{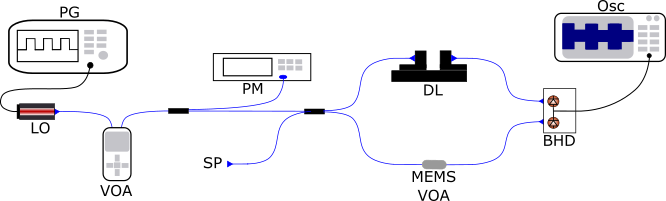}
 \caption{Schematics of the setup. The LO is pulsed at 50 MHz via gain switching. PG: pattern generator; LO: local oscillator; VOA: variable optical attenuator; PM: power meter; SP: signal port; DL: delay line; BHD: balanced homodyne detector; Osc: oscilloscope.}
 \label{setup}
\end{figure}

\noindent We now move on to show the phase-randomized SDI CV-QRNG in operation.
The setup is shown in Fig.~\ref{setup}.
The LO is a 1550 nm laser diode, with an integrated optical isolator, gain-switched to produce phase randomized pulses.
Its output first travels through a variable optical attenuator (VOA) and is then split by a 99:1 fibre coupler.
The 1\% output is connected to a power meter to monitor the power of the LO.
The 99\% output is split by a 50:50 coupler. The other input of the 50:50 coupler is left open such that any input state potentially controlled by an adversary could enter.
A microelectromechanical systems (MEMS) VOA on one output arm of the 50:50 coupler balances the power incident on the two photodiodes of a commercial wideband homodyne detector.
An optical delay line is used to match the arrival times of the pulses.
The output of the BHD is digitised using an oscilloscope with an analog-to-digital converter (ADC) resolution of eight bits and a sampling frequency of 40 GSamples/s.
The main advantage of this protocol is that the setup required is identical to a typical trusted CV-QRNG despite offering SDI assurance.
The phase randomization of the LO is a vital part of the security of this protocol.
In practical future implementations, in addition to the power meter for monitoring the intensity, Alice could add an interferometer to monitor the actual phase randomization of the LO.
The LO could be further protected from potential external phase seeding attacks by placing an additional optical isolator in front of it.

To gain-switch the laser, the dc bias is set just below threshold and the laser is driven above threshold by applying an ac voltage from a pattern generator.
When the laser cavity is empty, the lasing action is triggered entirely by spontaneous emission, which inherits its random phase from the vacuum \cite{abellan2014ultra,yuan2014robust}.
This condition holds for repetition frequencies up to 2.5 GHz \cite{yuan2014robust}.
However, we limit the clock rate to 50~MHz to minimise the signal ringing due to the imperfect response of the BHD circuit to higher frequency pulses.
\begin{figure}
\centering
\includegraphics[width=0.49\textwidth]{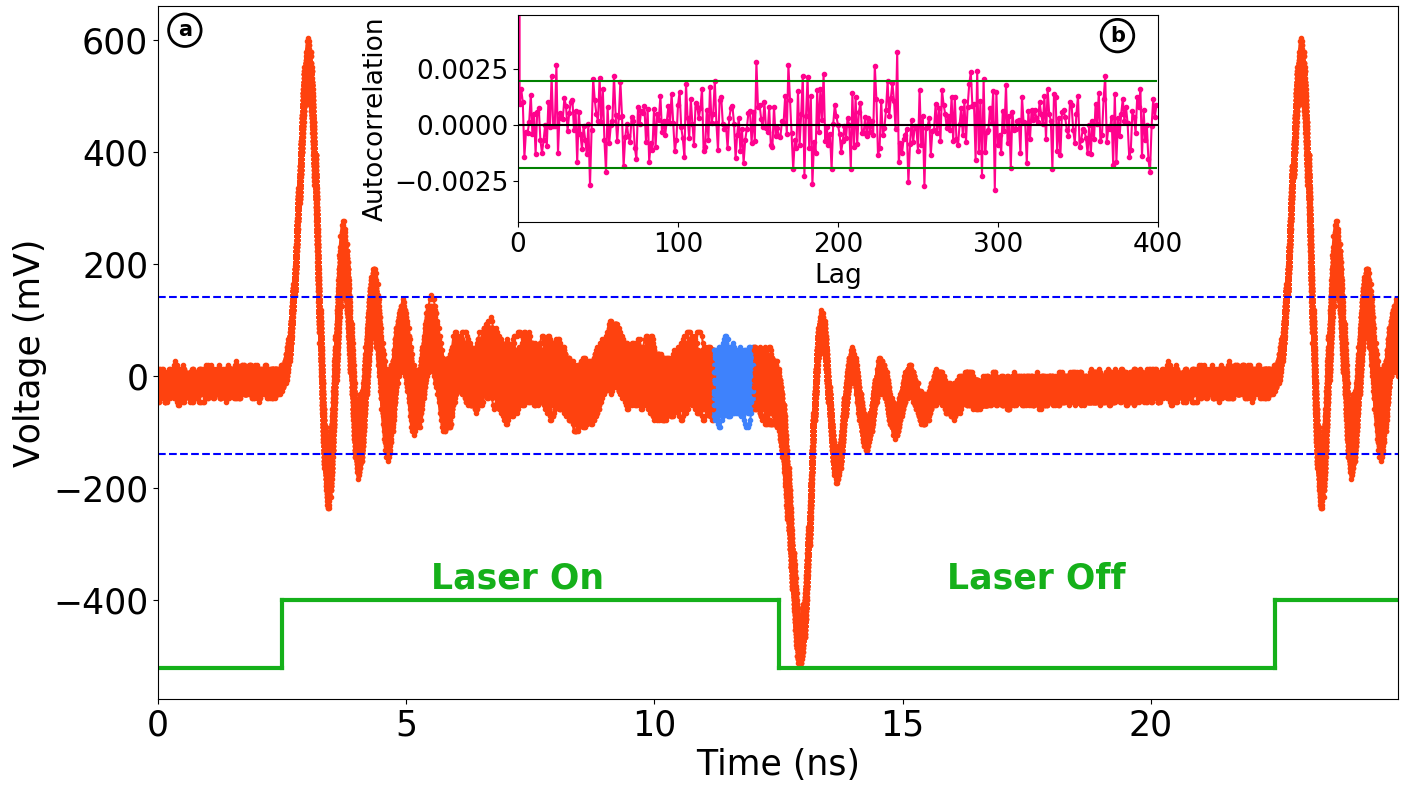}
 \caption{\textbf{(a)} Example of the ringing observed in the output of the BHD when the LO is pulsed at 50 MHz with a duty cycle of 50\%. The ac driving signal applied to the LO is shown in green, showing where the laser is on and off. The region from which samples were taken to generate the raw random numbers is highlighted in blue. The dashed lines show the ADC range used when acquiring data. \textbf{(b)} Autocorrelation evaluated on $10^6$ filtered raw data points with 95\% confidence intervals for a white noise process (green), showing that this data is uncorrelated. }
 \label{fig_ring}
\end{figure}

An example of the ringing observed is shown in Fig. \ref{fig_ring}A, in which the region from which the raw random numbers were sampled is highlighted.
The chosen pulsing frequency also allows us to minimize the correlations introduced by the finite bandwidth of the detector \cite{Shen2010practical}.

Filtering and randomness extraction are performed offline. We first apply a 1.6 GHz low pass filter to remove the noise above the bandwidth of the detector, then sub-sample the resulting data taking one point every laser pulse, giving an equivalent sampling rate of 50 MSamples/s.
The low frequency noise is removed by modulating at 25 MHz then applying a low pass filter.
The autocorrelation evaluated on a set of $10^6$ filtered points with the 95\% confidence intervals for lags of 0 to 400 is reported in Fig.~\ref{fig_ring}B, showing the absence of correlations due to low-frequency noise.

\section{Bounding the Min-Entropy}

\begin{figure}[t!]
\centering
\includegraphics[width=0.49\textwidth]{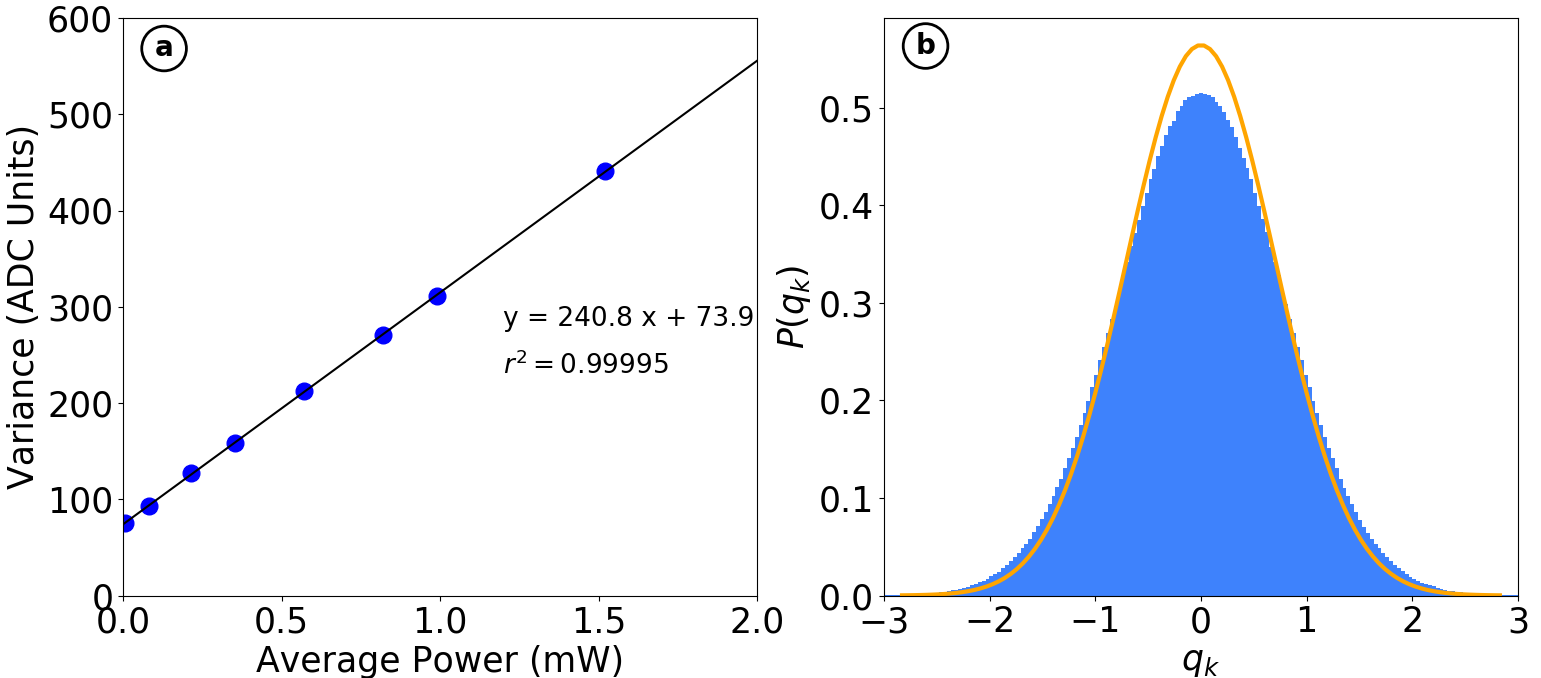}
 \caption{\textbf{(a)} Typical calibration line obtained during data acquisition, where the average power incident on each photodiode has been calculated from the power-meter measurements. \textbf{(b)} Probability density function (PDF) of filtered raw data converted into vacuum units (blue). Theoretical PDF for vacuum state input in the absence of excess noise (orange). }
 \label{fig_cal}
\end{figure}

To bound the conditional min-entropy, we estimate the resolution $\delta$ in vacuum units.
During our practical calibration, the signal port is blocked to provide a reference vacuum state input.
We measure the variance of the filtered data at different LO powers $P$ and fit a calibration line.
The intercept corresponds to the contribution of the electronic noise to the overall variance, whereas the gradient, $m$, can be used to estimate the contribution of the quantum noise.
A typical calibration line is shown in Fig.~\ref{fig_cal}A.
In the absence of electronic noise, the variance in ADC units would be given by $mP$ and the measurement resolution in vacuum units $\delta=\frac{\delta_{ADC}}{\sqrt{2mP}}$, where $\delta_{ADC}$ is the resolution of the oscilloscope ADC.
The solid line in Fig.~\ref{fig_cal}B represents the theoretical vacuum distribution used to bound the min-entropy of the raw numbers whose distribution is represented by the histogram.
According to our framework, Alice does not make any assumptions on the input state entering the signal port and therefore on the raw distribution that she will observe.
However, since in our proof of principle experiment there was no external source, it is reasonable to assume that the vacuum was actually the main input state.
The histogram of the raw data is then Gaussian but wider than reference vacuum distribution because it includes excess noise.

Using Eqs. (\ref{Entropy-PR-Bound}) and Eq. (\ref{bound_vacuum}), we obtain a typical conditional min-entropy of $H_{\text{min}} \left( Q_{\delta}|E\right)_{\text{pr}} \geq 5.53$ bits.
To extract iid bits we implement a Toeplitz hashing using a seed from another QRNG, described in \cite{marangon2018long}.
Given the length of the input string, the length of the seed was chosen to obtain a probability $\epsilon \geq 2^{-100}$ of distinguishing the output data distribution from a uniform one \cite{tomamichel2011leftover,Frauchiger2013}.
As a result, 5.4 random bits were distilled from each raw 8 bit sample.
With the 50 MHz sampling rate, this provides a secure generation rate of 270 Mbit/s.

To assess the implementation of the randomness extractor, we applied two standard statistical tests,  NIST \cite{nist,nistURL} and TestU01 \cite{testu01}.
The data gathered was split into blocks of 125 MB for the NIST tests.
The Rabbit and Alphabit batteries from the TestU01 suite were applied to all 900 MB of data at once.
The post-processed data passed all of these tests.
Detailed results  are reported in Appendix B.

\section{Conclusion}
In this work we presented an experimental SDI CV-QRNG based on phase randomized balanced homodyne detection capable of generating secure random numbers at an equivalent rate of 270 Mbit/s. Due to the SDI nature of the generator, no assumption on the input state was required.

The  achieved generation rate was limited by the ringing observed in the output of the balanced homodyne detector.
Any reduction of this impairment could significantly increase the generation rate.

In contrast to earlier SDI CV-QRNGs, this implementation does not require active optical components or the use of heterodyne detection. The gain-switched local oscillator provides the necessary phase randomization for the QRNG without adding components such as a phase randomizer and a random number generator to drive it. This also makes the setup robust against attacks probing the internal components. These features and the overall compactness of the generator are promising for a future integration on chip.

\begin{acknowledgments}
This project has received funding from the European Union’s Horizon 2020 research and innovation program under Marie Sklodowska-Curie Grant Agreement No. 750602,“Development of an Ultra-Fast, Integrated, Certified SecureQuantum Random Number Generator for Applications in Science and Information Technology”(UFICS-QRNG). P.R.S. gratefully acknowledges financial support from the EPSRC (Award No. 1771797) CDT in Integrated Photonic and Electronics Systems and Toshiba Research Europe, Limited.
\end{acknowledgments}

\section*{Appendix A: Equivalence between phase-randomized input and phase-randomized local oscillator}

In the following, we will explicitly demonstrate $\state E^{\text{I}} = \state E^{\text{II}} $, where $\state E^{\text{I}}$and $ \state E^{\text{II}} $ are defined in Eqs. (\ref{first_equivalence}) and (\ref{second_equivalence}) in the Main Text. We will argue that from a security perspective it is equivalent to place a phase randomiser at the input of the generator or to use a phase-randomized local oscillator. The equivalence will be proven by showing that Eve's reduced density matrix is the same in the two cases.

\begin{figure}[t!]
 \centering
  \includegraphics[width=0.49\textwidth]{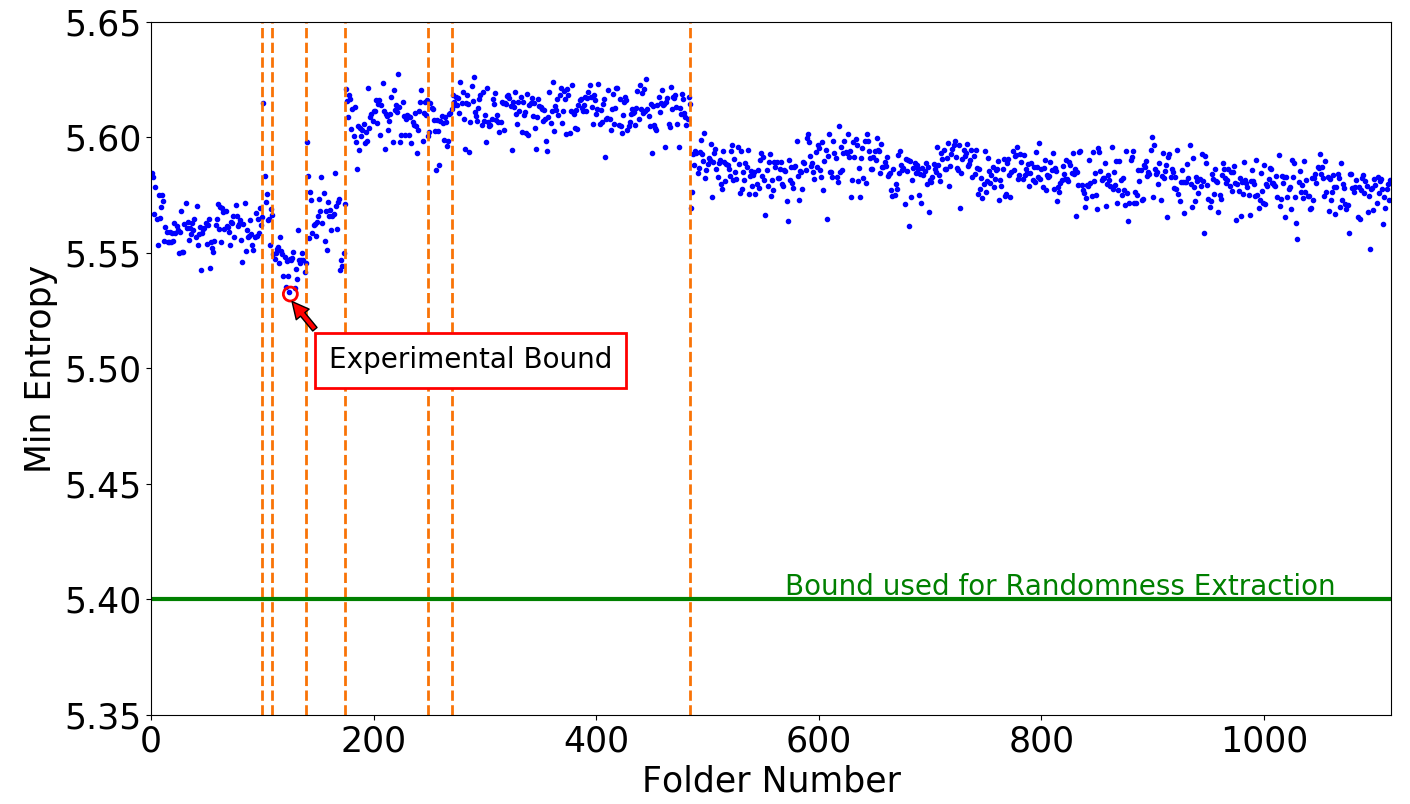}
 \caption{The blue points are the min-entropies corresponding to each data set acquired.
 The dashed lines indicate separate sessions in between which the setup was adjusted.
 For each session the entropy was estimated multiple times by periodically acquiring a calibration line approximately every 10 min. Hence, in a session multiple data sets were acquired, each of them with its own min-entropy bound. The minimum value of 5.53, circled in red, was used as the experimental bound for the min-entropy. Given the length of the input string, the length of the seed was chosen to obtain a probability $\epsilon \geq 2^{-100}$ of distinguishing the output data distribution from a uniform one. As indicated by the green horizontal line, we then distill 5.4 random bits from each raw 8-bit sample.}
 \label{experimental_entropies}
\end{figure}

\begin{table}[h!]
\begin{tabular}{l c c c}    \toprule
Statistical test & P value &  Proportion &  Result \\
\hline
Frequency                                & 0.156 & 0.990 & Success \\
Block Frequency                      &0.567& 	0.990& Success \\
Cumulative Sums                     &0.917 & 	0.984& Success \\
Cumulative Sums                     &0.038 & 0.991& Success \\
Runs                                         &0.512& 0.987& Success \\
Longest Run                             &0.668 & 	0.984& Success \\
Rank                                         &0.660 & 0.994 & Success \\
FFT                                           &0.445 & 0.985& Success \\
Non Overlapping Template       & 0.483 & 0.990	& Success \\
Overlapping Template              &0.777 & 0.989& Success\\
Universal                                  &0.101 & 0.987 & Success\\
Approximate Entropy              &0.145 & 0.992 & Success\\
Random Excursions                 & 0.384 & 0.991 & Success\\
Random Excursions Variant    & 0.335 & 0.992 & Success\\
Serial                                        &0.770 & 0.990 & Success\\
Serial                                        &0.724 & 0.991 & Success\\
Linear Complexity                   &0.714 & 0.989 & Success\\ 
\hline
\end{tabular}
\caption{Results of the NIST test battery applied on $10^3$ strings, each having a length of $10^6$ bits.}\label{nist_results}
 \label{table}
\end{table}

The most general Alice-Eve density matrix written in the Fock basis is
\begin{equation}\label{SharedState}
\state {AE} = \sum_{k,l,n,m} \rho^{k,l}_{n,m}\ket{k}_E\bra{l} \otimes \ket{n}_A\bra{m},
\end{equation}
where
$\left\{k_E\right\}_{k= 0\dots \infty}$ and $\left\{l_E\right\}_{l= 0\dots \infty}$
are Eve's basis states and
$\left\{n_A\right\}_{n= 0\dots \infty}$ and $\left\{m_A\right\}_{m= 0\dots \infty}$
are Alice's basis states.

We define the phase shift operator $\hat{U}_{\varphi} = e^{-i\varphi \hat{n}}$, where $\hat{n}$ is the photon number operator, and rewrite Eq. (4) of the main text as

\begin{align}\label{post-measu}
\state {AE}^{\text{pr}} &=  \sum_{k,l,n,m}\rho^{k,l}_{n,m} \ket{k}_E\bra{l} \otimes  \left( \frac{1}{2\pi}\int^{2\pi}_0 \text{d}\varphi\hat{U}_{\varphi} \ket{n}_A\bra{m}\hat{U}_{\varphi}^{\dagger} \right) \nonumber\\
& = \sum_{k,l,n,m}\rho^{k,l}_{n,m} \ket{k}_E\bra{l} \otimes  \left( \frac{1}{2\pi}\int^{2\pi}_0 \text{d}\varphi e^{{ -i(n-m)}\varphi} \ket{n}_A\bra{m} \right)  \nonumber\\
& = \sum_{k,l,n}\rho^{k,l}_{n,n}\ket{k}_E\bra{l} \otimes  \ket{n}_A\bra{n} \,.
\end{align}

We then consider the action of Alice's quadrature operator.
For ease of notation, in the following we will use the quadrature projector in the approximation of infinite resolution $\hat{Q}_\theta= \ketbra {q_\theta} {q_\theta}$, by dropping the reference to the interval $\delta$ and outcome $k$.

We then have
\begin{equation}\label{equivalence}
\left(\text{Id}_E\otimes\hat{Q}_{\theta}\right) \state {AE}^{\text{pr}} \left(\text{Id}_E\otimes\hat{Q}_{\theta}\right) ^{\dagger}
\end{equation}
and evaluate the reduced state of Eve referred to in the main text by $\state E^{I}$ by tracing out Alice's degrees of freedom:

\begin{align}\label{reduced1}
\state E^{I} & = \text{tr}_A \left[\left(\text{Id}_E\otimes\hat{Q}_{\theta}\right) \state {AE}^{\text{pr}} \left(\text{Id}_E\otimes\hat{Q}_{\theta}\right) ^{\dagger}\right] \nonumber\\
& = \sum_r \bra{r} \left[ \sum_{k,l,n}\rho^{k,l}_{n,n} \ket{k}_E\bra{l} \left( \hat{Q}_{\theta} \ket{n}_A\bra{n} \hat{Q}_{\theta}^{\dagger} \right) \right] \ket{r}\nonumber\\
& = \sum_{k,l,n}\rho^{k,l}_{n,n} \ket{k}_E\bra{l} \sum_r \braket{r} {{ q_{\theta}}} \braket{{ q_{\theta}}} {n}_A \braket{n} {{ q_{\theta}}} \braket{{ q_{\theta}}} {r}\nonumber\\
&=\sum_{k,l,n}\rho^{k,l}_{n,n} \ket{k}_E\bra{l} \: |\braket {{ q_{\theta}}} {n}_A|^2 \sum_r \braket { q_{\theta}} r \braket r { q_{\theta}}\nonumber\\
&=\sum_{k,l,n}\rho^{k,l}_{n,n} \ket{k}_E\bra{l} \:  |\braket {{ q_{\theta}}} {n}_A|^2 \bra{{ q_{\theta}}} \sum_r \ket r \braket r { q_{\theta}}\nonumber\\
&=\sum_{k,l,n}\rho^{k,l}_{n,n} \ket{k}_E\bra{l} \: |\braket {{ q_{\theta}}} {n}_A|^2
\end{align}

We now consider Alice applying a randomly phase-shifted quadrature operator $\hat{Q}^{\text{ps}}_{\theta,\phi}= \hat{U}_{\varphi}\hat{Q}_{\theta}\hat{U}^{\dag}_\varphi$ on her part of the system, such that now the overall phase averaged state is:
\begin{widetext}
\begin{align}\label{equivalence}
\state {AE}^{\text{pr}}
& =\int^{2\pi}_0 \frac{d\varphi}{2\pi}  \left(\text{Id}_E\otimes \hat{Q}^{\text{ps}}_{\theta,\phi}\right) \state {AE} \left(\text{Id}_E\otimes \hat{Q}^{\text{ps}}_{\theta,\phi} \right)^\dagger \nonumber\\
 &= \sum_{k,l,n,m}\rho^{k,l}_{n,m} \ket{k}_E\bra{l}  \otimes  \frac{1}{2\pi}\int^{2\pi}_0 d\varphi \left(\hat{U}_{\varphi}\hat{Q}_{\theta}\hat{U}^{\dag}_\varphi \right) { \ket  {n}_A \bra{m}} \left( \hat{U}_{\varphi}\hat{Q}_{\theta}\hat{U}^{\dag}_\varphi \right) \nonumber\\
&=  \sum_{k,l,n,m}\rho^{k,l}_{n,m} \ket{k}_E\bra{l}  \otimes \frac{1}{2\pi}\int^{2\pi}_0 d\varphi e^{-i(m-n)\varphi}\hat{U}_{\varphi}\hat{Q}_{\theta} { \ket  {n}_A \bra{m}} \hat{Q}_{\theta} \hat{U}_{\varphi}^{\dagger}
\end{align}
\end{widetext}
\begin{widetext}
By tracing out Alice's degrees of freedom, we obtain Eve's density matrix $\state E^{II}$:
\begin{align}\label{reduced2}
\state E^{II} &= \text{tr}_A \left[\state {AE}^{\text{pr}} \right] \nonumber\\
&= \sum_r \bra r \left( \sum_{k,l,n,m}\rho^{k,l}_{n,m} \ket{k}_E\bra{l} \otimes \frac{1}{2\pi}\int^{2\pi}_0 d\varphi  e^{-i(m-n)\varphi}\hat{U}_{\varphi}\ket { q_{\theta}} \braket { q_{\theta}} {{ n}}_A \braket{ m} {{ q_{\theta}}} \bra {{ q_{\theta}}} \hat{U}_{\varphi}^{\dagger} \right) \ket r\nonumber\\
&= \sum_{k,l,n,m}\rho^{k,l}_{n,m} \ket{k}_E\bra{l} \otimes \frac{1}{2\pi}\int^{2\pi}_0 d\varphi  e^{-i(m-n)\varphi}\sum_r \bra r\hat{U}_{\varphi}\ket { q_{\theta}} \braket { q_{\theta}} {n}_A \braket{m} {{ q_{\theta}}} \bra {{ q_{\theta}}} \hat{U}_{\varphi}^{\dagger} \ket r\nonumber\\
&= \sum_{k,l,n,m}\rho^{k,l}_{n,m} \ket{k}_E\bra{l} \otimes \frac{1}{2\pi}\int^{2\pi}_0 d\varphi  e^{-i(m-n)\varphi}\sum_{r,s}   \bra r\hat{U}_{\varphi}\ket s \braket s { q_{\theta}} \braket { q_{\theta}} {n}_A \braket{m} {{ q_{\theta}}} \bra {{ q_{\theta}}} \hat{U}_{\varphi}^{\dagger} \ket r\nonumber\\
&=\sum_{k,l,n,m}\rho^{k,l}_{n,m} \ket{k}_E\bra{l} \otimes   \frac{1}{2\pi}\int^{2\pi}_0 d\varphi e^{-i(m-n)\varphi}\sum_{r,s} e^{{ -i(s-r)}\varphi}\braket r s \braket s { q_{\theta}} \braket { q_{\theta}} {n}_A \braket{m} {{ q_{\theta}}} \braket {{ q_{\theta}}}  r\nonumber\\
&=  \sum_{k,l,n,m}\rho^{k,l}_{n,m} \ket{k}_E\bra{l} \otimes \frac{1}{2\pi}\int^{2\pi}_0 d\varphi e^{-i(m-n)\varphi} \sum_r \braket r { q_{\theta}} \braket { q_{\theta}} {n}_A \braket{m} {{ q_{\theta}}} \braket {{ q_{\theta}}}  r\nonumber\\
&= \sum_{k,l,n}\rho^{k,l}_{n,n} \ket{k}_E\bra{l} \:|\braket{{ q_{\theta}}}{n}_A|^2,
\end{align}
which is equal to Eve's density matrix in Eq.~(\ref{reduced1}), thus completing the proof.
\end{widetext}

\section*{Appendix B: Experimental bound to the min-entropy}
As explained in the main text, we calculate a bound on the min-entropy based on the gradient of a calibration line obtained by varying the power of the LO and measuring the variance of the filtered output. We assume that this relationship holds for the data gathered following this calibration. The performance of the system and hence the min-entropy is likely to change over time due to degradation of the components and changing environmental conditions. Our system therefore automatically obtains a new calibration line periodically (approximately every 10 min), allowing the value of the min-entropy used in the randomness extraction to be updated if necessary.
By taking into account the error in the gradient $m$ associated with the fit, we calculate conservative estimates of the min-entropy from the calibration lines obtained when gathering the data discussed in the main text.
The resulting values are plotted in Fig.~\ref{experimental_entropies}.
The vertical dashed lines indicate when parts of the setup were adjusted, changing the maximum LO power incident on the detector.
As expected, we see a corresponding change in the min-entropy. This highlights our systems’ ability to respond to changes in operating conditions and continue to extract iid bits.
The difference between the largest and smallest values of min-entropy obtained over all of the acquisitions is less than 2 \%. The corresponding difference over the longest uninterrupted set of acquisitions is less than 1 \%, highlighting the stability of our system. Furthermore, the number of iid bits extracted from each 8 bit sample, shown in green, is far below the minimum min-entropy bound obtained compared to the variation in values seen.
\section*{Appendix C: Result of the NIST tests}
In Table \ref{table}, the results of a typical run of the NIST test  are reported.
The test is applied on $10^3$ strings after application of the randomness extractor, and each string has a length of $10^6$ bits.

\end{document}